\documentclass[twocolumn,
amsmath,amssymb,prb]{revtex4-1}

\usepackage{graphicx}
\usepackage{bm}
\usepackage{tikz}
\usetikzlibrary{decorations.pathmorphing}
\usepackage{floatrow}
\usepackage{multirow}
\usepackage{tabularx}
\usepackage{color}

\usepackage{comment}
\usepackage{color}
\usepackage[colorlinks,urlcolor=blue,citecolor=blue,linkcolor=blue]{hyperref}
\usepackage{cleveref}
\usepackage{physics}
\usepackage{mathtools}
\usepackage{mathrsfs}
\usepackage{booktabs}

\newcommand{\sch}{Schr{\"o}dinger }
\newcommand{\w}{\omega}
\newcommand{\be}{\begin{equation}}
\newcommand{\ee}{\end{equation}}

\newcommand{\ch}{\hat{c}}
\newcommand{\chd}{\hat{c}^{\dagger}}
\newcommand{\eh}{\hat{e}}
\newcommand{\ehd}{\hat{e}^{\dagger}}

\newcommand{\iu}{{i\mkern1mu}}
\newcommand{\vk}{\vectorbold{k}}
\newcommand{\vp}{\vectorbold{p}}
\newcommand{\vq}{\vectorbold{q}}
\newcommand{\vkd}{\vectorbold{k}^{'}}
\newcommand{\e}{\epsilon}

\begin{document}

\title{Microscopic theory of excitons bound by light}
\author{Sangeet S. Kumar}
\affiliation{ARC Centre of Excellence in Future Low-Energy Electronics Technologies and School of Physics and Astronomy, Monash University, Victoria 3800, Australia}
\author{Meera M. Parish} 
\affiliation{ARC Centre of Excellence in Future Low-Energy Electronics Technologies and School of Physics and Astronomy, Monash University, Victoria 3800, Australia}
\author{Jesper Levinsen}
\affiliation{ARC Centre of Excellence in Future Low-Energy Electronics Technologies and School of Physics and Astronomy, Monash University, Victoria 3800, Australia}

\date{\today}

\begin{abstract}

We theoretically investigate the scenario of a semiconductor quantum well in a microcavity, where the band structure is arranged such that optically excited electron-hole pairs cannot form Coulomb-bound excitonic states. 
However, it is still possible to form exciton polaritons (part-light, part-matter quasiparticles), where the excitons are bound via the exchange of microcavity photons rather than via Coulomb interactions. Using a diagrammatic theory, we determine the spectral response of the semiconductor microcavity, which includes exciton-polariton resonances as well as a continuum of unbound electron-hole pairs. Our method also gives us access to the photon fraction and the electron-hole wave function of the exciton polariton. In particular, we obtain the conditions under which an exciton is bound by photon exchange and we show that Coulomb interactions can enhance binding at large cavity photon frequencies. Our results for the spectral response are in good agreement with a recent experiment on doped quantum wells [{\rm E. Cortese et al.}, Nat. Phys. \textbf{17}, 31 (2021)]. 
\end{abstract}

\maketitle

\section{Introduction}

When there is a strong coupling between light and matter in a semiconductor microcavity, new hybrid quasiparticles called exciton polaritons emerge, which are superpositions of excitons (bound electron-hole pairs) and photons~\cite{LaussyFabriceP2017M,CarusottoCiutiReview2013,KeelingJ2007Ccip}. Here, the coherent energy exchange between an exciton and cavity photon is faster than the rate of dissipation and decoherence, such that light and matter become entwined to the point where they are inseparable.  
Exciton polaritons have enabled a range of many-body coherent quantum phenomena to be realized, such as  Bose-Einstein condensation~\cite{Kasprzak2006} and superfluidity~\cite{Amo2009c}. Most of these phenomena have been successfully described using a coupled oscillator model of the exciton polariton, where the internal structure of the exciton is neglected. However, recently it has been experimentally demonstrated~\cite{VeryStrong2} that excitons can be non-perturbatively modified in the so-called very strong coupling regime~\cite{VeryStrongCoupling,Citrin2003,MicroscopicJM,Laird2022}, where the light-matter coupling strength approaches the exciton binding energy. This raises the intriguing possibility of precisely manipulating the electron-hole wave function via the coupling to light, with potential applications in optoelectronics~\cite{Kockum2019,Forn-Diaz2019}.

Remarkably, in a recent experiment~\cite{NatureExBoundByLight}, Cortese \textit{et al.}~observed the formation of exciton polaritons in a semiconductor microcavity that did not support Coulomb-bound electron-hole pairs. The quantum well in this experiment was carefully engineered to feature an effective electron-hole Coulomb repulsion, and therefore the observed electron-hole binding could instead be attributed to an attractive potential originating from emission and reabsorption of cavity photons, as originally proposed in Ref.~\cite{StrongCouplingOfIonTrans}. The experimental results thus clearly defy an explanation in terms of coupled oscillators, as there is \textit{a priori} no matter quasiparticle for the photon to coherently couple to. 

Here, we use a microscopic approach that explicitly includes the electron, hole, and photon degrees of freedom to model excitons bound by microcavity photons. Our approach uses Feynman diagrams, and is based on the recent microscopic description of exciton polaritons in Ref.~\cite{MicroscopicJM}. It allows us to calculate the light-matter coupled state numerically exactly within the approximation of an inert Fermi sea of holes, which may be viewed as a variational ansatz similar to approaches in quantum impurity problems~\cite{Scazza2022}. Our calculation gives us access to the microcavity spectral function, including both discrete and continuous parts of the spectrum, which we find compares well with the experimental results~\cite{NatureExBoundByLight}. Furthermore, we obtain analytic expressions for the spectral function and electron-hole wave function under the assumption that Coulomb interactions can be neglected compared with the light-matter coupling strength, and we find that these closely match the numerically exact results in the regime of experimentally relevant parameters. Finally, our method allows us to obtain the conditions under which a exciton state bound by light exists in the semiconductor microcavity. 

The paper is organized as follows; the conceptual framework of the quantum well microcavity and the associated microscopic Hamiltonian is introduced in Section~\ref{Model}. In Section~\ref{sec:Results} we then calculate the spectral response of the microcavity, analyze the photon fraction and electron-hole wave functions contributing to the polariton state, and compare our results with experiment.
We conclude in Section~\ref{sec:conc}.

\section{Model} \label{Model}

\begin{figure}
\begin{center}
\includegraphics[width=0.95\linewidth]{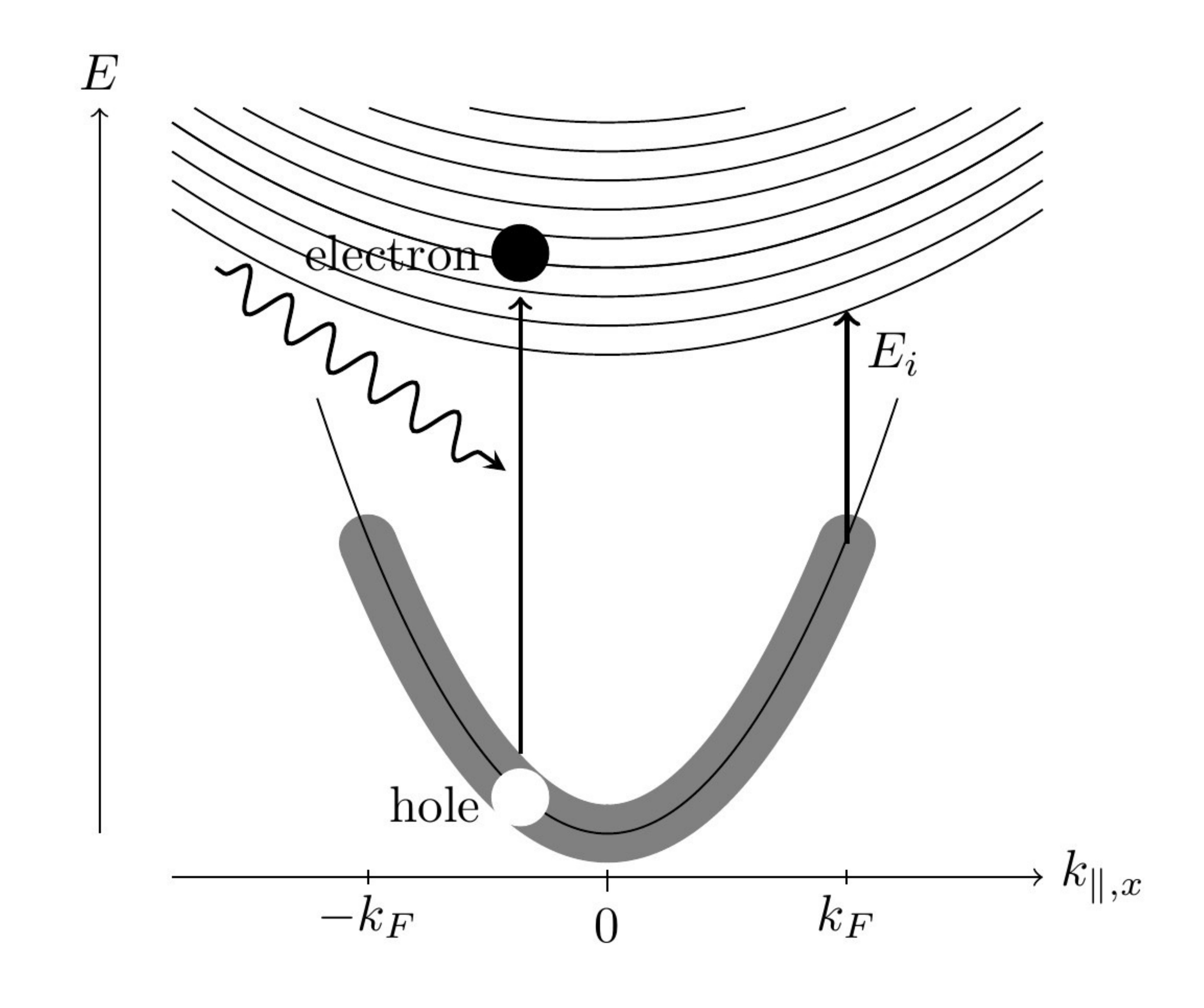}
\caption[system]{\label{QWell} 
Schematic of the band structure in the electron-doped quantum well microcavity that was used to observe excitons bound by photon exchange~\cite{NatureExBoundByLight}. The quantum well was designed to support a single bound electronic subband, which was filled (shaded gray) up to the Fermi momentum $k_F$ in the 2D plane. A photon (wavy line) can ionize an electron from the bound subband into the 3D unbound continuum (series of curves), thus resulting in an electron-hole pair. Since, along the 2D plane, the effective electron mass is greater in the ionization band than in the bound subband, the minimum ionization energy $E_i$ is on the electron Fermi surface.
}
\end{center}
\end{figure}

We consider the scenario investigated in the recent experiment by Cortese \textit{et al.}~\cite{NatureExBoundByLight},
namely a quantum well that supports a single bound electronic subband and no excitonic bound states. To model the experiment, we employ the band structure illustrated in Fig.~\ref{QWell}. It contains a single bound two-dimensional (2D) electronic subband --- with a positive effective mass $m_\mathrm{sub}$ --- which is occupied up to the Fermi momentum $k_F$ as represented by the grey shaded region. An electron from the bound subband can be ionized into an unbound 3D continuum, represented by a series of parallel curves, with in-plane effective mass $m_{e\parallel}$. The resulting in-plane hole has effective mass $m_{h\parallel}=-m_\mathrm{sub}$ under the usual electron-hole mapping. Importantly, $|m_{h\parallel}|<m_{e\parallel}$ and therefore the total in-plane kinetic energy of the electron-hole pair is negative, which can be seen schematically in Fig.~\ref{QWell} from the shrinking ionization energy with increasing momentum.
Accordingly, the kinetic energy has the same sign as the attractive electron-hole Coulomb interaction, i.e., the situation is formally analogous to two electrons interacting via a repulsive Coulomb potential. This effective repulsion is what prevents conventional exciton formation.

To describe the quantum well microcavity system, we use a Hamiltonian that consists of contributions from the light, matter, and light-matter coupling terms. The matter term $\hat{H}_m$ describing the band structure illustrated in Fig.~\ref{QWell} is
\be
\begin{split}\label{eq:Ham}
\hat{H}_{m} 
 =&  \underset{\vk}{\sum} \bigg{[} 
( \e_{e \vk}+E_g/2)\, \ehd_{\vk} \eh_{\vk} +
( \e_{h \vk}+E_g/2)\, \hat{h}^{\dagger}_{\vk} \hat{h}_{\vk} \bigg{]} \\ 
& +\frac{1}{2} \underset{\vk \vkd \vq}{\sum} V(\vq) \bigg[
\ehd_{\vk+\vq} \ehd_{\vkd-\vq} \eh_{\vkd} \eh_{\vk} +
\hat{h}^{\dagger}_{\vk+\vq} \hat{h}^{\dagger}_{\vkd-\vq} \hat{h}_{\vkd} \hat{h}_{\vk}  \\
& \qquad \qquad -2 \ehd_{\vk+\vq}\hat{h}^{\dagger}_{\vkd-\vq}\hat{h}_{\vkd}\eh_{\vk}
\bigg] .
\end{split}
\ee
Here, and in the following, we work in units where the volume and $\hbar$ are both set to unity.
In Eq.~\eqref{eq:Ham}, the creation (annihilation) operators of electrons and holes at momentum $\vk$ are written as $\ehd_{\vk} \text{ } (\hat{e}_{\vk})$ and $\hat{h}^\dagger_{\vk} \text{ } (\hat{h}_{\vk})$, respectively. To account for the dimensionality of the system, we separate the momenta and effective masses into the quantum well plane and the transverse direction labelled by $\parallel$ and $z$, respectively. The total kinetic energies of the electron and the holes become $\e_{e \vk}=\frac{\vk_\parallel^2}{2m_{e \parallel}}+\frac{k_z^2}{2m_{e z}}$ and $\e_{h \vk}=\frac{\vk_\parallel^2}{2m_{h \parallel}}$, where we have accounted for the 2D nature of the bound subband by taking the transverse hole effective mass to be infinite. We furthermore take the transverse electron effective mass $m_{ez}$ to be the bulk GaAs electron mass $m_{e_{GaAs}}$ (see Table~\ref{ParameterTable} for a summary of theory parameters), while the in-plane effective masses $m_{e\parallel}$ and $m_{h\parallel}$, which are specific to the quantum well, are treated as free parameters. 

The kinetic terms in Eq.~\eqref{eq:Ham} additionally contain the band gap $E_g$ defined as the energy difference between the bound subband and the ionization band at zero momentum. However, as illustrated in Fig.~\ref{QWell}, the actual minimum ionization energy $E_i$ is obtained for excitations at the Fermi surface:
\begin{align} \label{eq:EiEg}
    E_i=E_g+\frac{k_F^2}{2m_{e \parallel}}-\frac{k_F^2}{2m_{h \parallel}}
    = E_g-\frac{k_F^2}{2\mu_\parallel},
\end{align}
where we define the (positive) mass $\mu_\parallel=1/(m_{h\parallel}^{-1}-m_{e\parallel}^{-1})$. Importantly, our theory described below only depends on $\mu_\parallel$ and not $m_{e\parallel}$ and $m_{h\parallel}$ separately. 

To describe the electronic interactions, in Eq.~\eqref{eq:Ham} we use the 3D Coulomb potential $V(\vq)=\frac{1}{2m_r a_0} \frac{8\pi}{q^2}$ which is written in terms of the GaAs reduced effective mass $m_r=(m_{e_{GaAs}}^{-1}+m_{h_{GaAs}}^{-1})^{-1}$ and the Bohr radius $a_0 \approx 16{\rm nm}$. These parameters should produce a realistic Coulomb interaction strength, since they yield the expected 2D exciton binding energy of $\approx 10$ meV in a standard GaAs quantum well with Coulomb-bound exciton states~\cite{ExBindingEnergy}. 

We now turn to the theoretical description of the resonant cavity photon mode. For simplicity we consider only photons at normal incidence. Therefore, the microcavity photon is described by
\be
\hat{H}_p = \w  \chd \ch .
\ee
Here, the operators $\chd$ and $\ch$ create and annihilate a microcavity photon, respectively, with the resonant cavity photon energy $\w$.

Finally, the light-matter component $\hat{H}_g$ describes the transformation of a microcavity photon to an electron-hole pair and vice versa:
\be \label{eq:Hg}
\hat{H}_g 
= g \underset{\vk_\parallel,k_z,k_z'}{\sum} \chi(k_z') \, \ehd_{\vk_\parallel,k_z} \hat{h}^{\dagger}_{-\vk_\parallel,k_z'} \ch +
h.c. 
\ee
Here we apply the rotating wave approximation, which is justified since we are working with photon energies and resonator energies $\w$ which are near resonant with the large band gap. We furthermore use a momentum independent constant $g$ to couple the photon and electron-hole states, where $\chi(k_z)$ is the envelope function for the hole that is tightly bound along the transverse $z$ direction, with normalization $\sum_{k_z}|\chi(k_z)|^2 =1$. Note that conservation of momentum requires that the electron-hole pair has zero total in-plane momentum when it is generated by a photon at normal incidence.

The total Hamiltonian that describes matter, light, and light-matter coupling terms is thus
\begin{align}
    \hat H=\hat H_m+\hat H_p+\hat H_g.
\end{align}
Formally, this is similar to the Hamiltonian employed in Ref.~\cite{MicroscopicJM}, which performed a microscopic calculation of exciton polaritons in a microcavity that exactly incorporated the light-induced changes of the electron-hole wave function due to strong light-matter coupling. In that work it was argued that the resonant cavity photon frequency should be renormalized due to the use of contact light-matter interactions, as in Eq.~\eqref{eq:Hg} (see also Refs.~\cite{Hu2020QuantumFI,Li2021PRB}). However, in the present work this is not an issue since the bound subband can only be excited at momenta up to $k_F$, and hence the electron-hole wave function is not probed at arbitrarily short length scales.

\begin{table}[h]
\begin{tabular}{ |c|c|c|c| } 
\hline
\multicolumn{4}{|c|}{Physical parameters}\\
\hline
\multicolumn{2}{|c|}{Experimental parameters} & \multicolumn{2}{|c|}{
Theoretical inputs}\\
\hline
\multicolumn{1}{|p{1.9cm}|}{ 
\centering $E_i$} 
& \multicolumn{1}{|c|}{138meV} 
& \multicolumn{1}{|p{2cm}|}{
\centering $g$} 
& \multicolumn{1}{|c|}{${\rm 360meVnm^{3/2}}$ }\\

\multicolumn{1}{|p{1.9cm}|}{
\centering $n$} 
& \multicolumn{1}{|c|}{$5\times 10^{12} {\rm cm}^{-2}$} 
& \multicolumn{1}{|p{1.9cm}|}{
\centering $g^*$} 
& \multicolumn{1}{|c|}{${\rm 360meVnm^{3/2}}$ }\\

\multicolumn{1}{|p{1.9cm}|}{
\centering $m_{e_{GaAs}}$} 
& \multicolumn{1}{|c|}{$0.063m_0$} 
& \multicolumn{1}{|p{1.9cm}|}{
\centering $\mu_\parallel$} 
& \multicolumn{1}{|c|}{$0.63 m_0$ }\\

\multicolumn{1}{|p{1.9cm}|}{
\centering $m_{h_{GaAs}}$} 
& \multicolumn{1}{|c|}{$0.51m_0$} 
& \multicolumn{1}{|p{1.9cm}|}{
\centering $\mu_\parallel{^*}$} 
& \multicolumn{1}{|c|}{$0.11 m_0$ }\\

\multicolumn{1}{|p{1.9cm}|}{
\centering $m_{e z}$} 
& \multicolumn{1}{|c|}{$m_{e_{GaAs}}$} 
& \multicolumn{1}{|p{1.9cm}|}{
\centering $\Gamma$} 
& \multicolumn{1}{|c|}{$5 {\rm meV}$}\\

\multicolumn{1}{|p{1.9cm}|}{
\centering $a_0$} 
& \multicolumn{1}{|c|}{$16{\rm nm}$} 
&
\multicolumn{1}{|p{1.9cm}|}{
\centering $\Gamma_{eh} $} 
& \multicolumn{1}{|c|}{$2{\rm 
meV}$}\\

\hline
\end{tabular}
\caption{\label{ParameterTable}
Table of parameters used for our modelling of the experiment in Ref.~\cite{NatureExBoundByLight}. Here we take the ionization energy $E_i$ and 2D electron density $n$ from the experiment. The Fermi momentum $k_F$ is then obtained from $n$ using the 2D ideal Fermi gas expression, $k_F=\sqrt{2\pi n}$. GaAs effective masses for the electron $m_{e_{GaAs}}$ and hole $m_{h_{GaAs}}$ are taken to be standard values in the literature~\cite{levinshtein1996gallium} in terms of $m_0$ the vacuum electron mass. The Bohr radius $a_0$ used in the Coulomb potential yields the expected 2D exciton binding energy ($\approx 10$ meV) in a standard GaAs quantum well~\cite{ExBindingEnergy}. On the right, we have the parameters specifically chosen in this work: the light-matter coupling constant for the case with and without the Coulomb interaction, $g^*$ and $g$, respectively, and the in-plane reduced effective mass of the electron-hole pair for the case with and without the Coulomb interaction, $\mu_\parallel^*$ and $\mu_\parallel$, respectively. The photon linewidth $\Gamma$ is independent of whether we include Coulomb interactions, and we include an electron-hole linewidth parameter $\Gamma_{eh}$ for convergence of our numerical results.
}
\end{table}

\section{Excitons bound by strong coupling to microcavity photons} \label{sec:Results}

We now discuss the formation of exciton polaritons within our model Hamiltonian. 
In particular, we use a variational approach to determine the photon amplitude and electron-hole wave functions, allowing us to investigate in detail the binding of electron-hole pairs. The photon amplitude furthermore allows us to calculate the spectral response of the system through the spectral function, which we find agrees well with the experiment~\cite{NatureExBoundByLight}. 

\subsection{Formalism} \label{Solution}
To describe the bound state of an electron-hole pair due to the strong coupling to light, we consider the state
\begin{align}\label{eq:state}
\ket{\Psi}= \underset{\vk_\parallel,k_z,k_z'}{\sum} \chi(k_z') \phi({{\vk_\parallel},k_z}) \ehd_{{\vk_\parallel},k_z} \hat{h}^{\dagger}_{{-\vk_\parallel},k_z'} \ket{0}+\gamma\hat c^\dag\ket{0},
\end{align}
which is a superposition of the electron-hole and the photonic components, respectively. Here, the hole is bound along the transverse $z$ direction, with envelope function $\chi(k_z')$ like in Eq.~\eqref{eq:Hg}. The electron-hole wave function $\phi$ and the photon amplitude $\gamma$ satisfy the normalization condition $\underset{ \vk_\parallel, k_z}{\sum} |\phi(\vk_\parallel, k_z)|^2+|\gamma|^2=1$. Note that in Eq.~\eqref{eq:state} we treat the Fermi sea as inert, i.e., we ignore the possibility of creating excitations within the bound subband electron Fermi sea. Such processes may be expected to be suppressed when there are no bound states associated with particles excited out of the Fermi sea. Since we consider a truncated Hilbert space, the state in Eq.~\eqref{eq:state} may be viewed as a variational ansatz for the ground state of the system~\cite{Parish2016}.

Projecting the \sch equation $\hat{H}\ket{\Psi}=E\ket{\Psi}$ onto the photon and electron-hole subspaces individually, we obtain the coupled set of equations for the wave function and energy $E$
\begin{subequations} \label{CoupledEqns}
\begin{align}
\big( E- E_g
-E_{\vk_\parallel,k_z}
\big) 
\phi (\vk_\parallel, k_z) &= \nonumber \\ & \hspace{-35mm} - \underset{\vq_\parallel,q_z}{\sum} n_{\vq_\parallel} V_s (\vk_\parallel-\vq_\parallel,k_z-q_z)  \phi (\vq_\parallel, q_{z})  +g\gamma,
\label{Eq1Coupled}\\
( E-\w ) \gamma  
&= g \underset{\vq_\parallel,q_z}{\sum}
n_{\vq_\parallel}
\phi (\vq_\parallel, q_{z}).
\label{Eq2Coupled}
\end{align}
\end{subequations}
In the first of these equations, we have replaced the Coulomb potential by its planar $s$-wave projection, since we are working with photons that only couple to electron-hole pairs in a relative $s$-state: 
\be
\begin{split}
&V_s(\vp-\vk
)= \int_0^{2\pi} \frac{d\varphi}{2\pi}\, V(\vp_\parallel-\vk_\parallel,p_{z}-k_{z}) \\
&= \frac{ 8 \pi 
/ (2m_r a_0)}{\sqrt{\left(p^2_\parallel-k^2_\parallel\right)^2+2 (p^{2}_\parallel+k^2_\parallel)(p_{z}-k_{z})^2+(p_{z}-k_{z})^4}},
\end{split}
\ee
where $\varphi$ is the in-plane angle between the relative momentum of incoming and outgoing electron-hole pairs, $\vk$ and $\vp$, respectively.
We have also defined the total kinetic energy of the electron-hole pair as $E_{\vk_\parallel,k_z}=-\frac{\vk_\parallel^2}{2\mu_\parallel}+\frac{k_z^2}{2m_{ez}}$, and introduced the Fermi-Dirac distribution of the holes in the quantum well plane, $n_{\vk_\parallel}$. In the following, we take zero temperature such that $n_{\vk_\parallel}=\Theta(k_F-|\vk_\parallel|)$ with $\Theta$ the unit step function; however our approach is straightforward to generalize to finite temperature.
We solve the set of equations in \eqref{CoupledEqns} numerically by introducing an appropriate momentum grid and treating it as an eigenvalue problem with eigenvalue $E$~\cite{NumericRec}. This approach yields a (discrete) set of photon fractions $\gamma_n$, electron-hole wave functions $\phi_n$, and energies $E_n$.

\begin{figure}[th]
\includegraphics[width=1.0\linewidth]{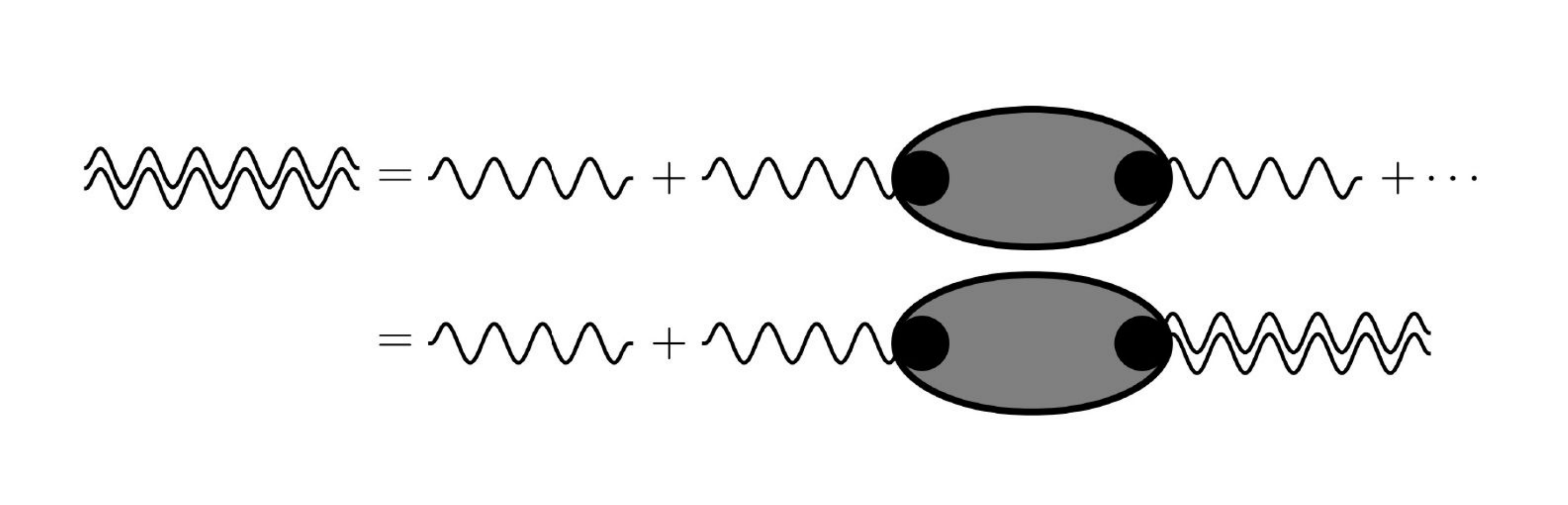}
\caption[system]{\label{DysonDiagram}
Diagrammatic representation of the Dyson equation~\eqref{Gc} satisfied by the dressed cavity photon propagator (double wavy line). The free photon propagator ${G}_c^{(0)} (E)$ is drawn with a single wavy line. The photon self energy $\Sigma (E)$ consists of the light-matter coupling $g$ (drawn as a black dot) and the full electron-hole propagator (shaded ellipse).   
}
\end{figure}

The spectral response of the quantum well microcavity is encoded in the spectral function~\cite{CiutiPRA2006,Cwik2016}
\begin{align}
A(E)=-\frac{1}{\pi} {\rm Im} [G_c(E)] ,    
\end{align}
with dressed cavity photon propagator:
\begin{align}
    {G}_c(E)=\expval*{ \hat{c}\frac{1}{E-\hat{H} +\iu \Gamma} \hat{c}^\dag}{0},
\end{align}
where we have introduced $\iu \Gamma $ as the photon linewidth parameter, which also serves to shift the pole structure of the propagator to the lower half of the complex plane. As discussed in Ref.~\cite{MicroscopicJM}, this in turn satisfies the Dyson equation illustrated in Fig.~\ref{DysonDiagram}:
\begin{align} \label{Gc}
{G}_c(E) & = G_c^{(0)}(E) + G_c^{(0)}(E) \Sigma(E) G_c(E) .
\end{align}
Here $G_c^{(0)}(E) = 1/(E-\omega +i \Gamma)$ is the free photon propagator, i.e., in the absence of light-matter coupling, while the self-energy from the coupling to matter is
\begin{align}
    \Sigma(E) = \expval*{ \hat{c} \hat{H}_{g} \frac{1}{E-\hat{H}_m+\iu \Gamma_{eh}} \hat{H}_{g} \hat{c}^\dag}{0} .
\end{align}
For a detailed derivation, see Appendix~\ref{AppendixA}. Note that we have introduced a separate linewidth $\Gamma_{eh}$ for the electron-hole part to account for any dissipation in the quantum well and to aid the numerical calculations. 

Formally, the \sch equation in Eq.~\eqref{CoupledEqns} provides an approximate treatment of the self energy within the Dyson equation, and its solution can be employed to yield the discrete version of the dressed photon propagator 
\be
\begin{split} \label{Discrete G}
& {G}_c (E)=\underset{n}{\sum} \frac{ \abs{\gamma_n}^2}{E-E_n+\iu \abs{\gamma_n}^2\Gamma+i(1-|\gamma_n|^2)\Gamma_{eh}} .
\end{split}
\ee
In general, we will assume that both $\Gamma$ and $\Gamma_{eh}$ are small such that they only negligibly affect the analytic results that we derive, i.e., we will effectively take them to vanish in all analytic expressions below.

A useful approximation is to assume that the strength of the light-matter coupling is such that the Coulomb interaction only provides a small correction to the results. In particular, completely neglecting the Coulomb interaction allows us to calculate $G_c (E)$ analytically by solving Eq.~\eqref{CoupledEqns} for the photon amplitude $\gamma$:
\begin{align}
G_c (E)&=\bigg( E-\w-\!g^2
\sum_{\vk,k_z}
\frac{n_{\vk_\parallel}}{E+\iu 0 -E_g - E_{\vk_\parallel,k_z} } \!+\!\iu 0 
\! \bigg)^{-1}\nonumber\\
&= \bigg(E-\w- 
\frac{ g^2 \mu_{\parallel} \sqrt{m_{ez}} }{ \sqrt{2} \pi }
\bigg(
\sqrt{E_i-E-\iu 0}-\nonumber\\  
&\hspace{25mm}\sqrt{E_g  -E -\iu 0}
\bigg) + \iu 0 
\bigg)^{-1}.\label{AnalyticProp}
\end{align}
This is a key result of this work, and it allows us to gain insight into the spectrum of the microcavity, as discussed below in Sec.~\ref{sec:resultsB}. In particular, Eq.~\eqref{AnalyticProp} explicitly shows how the coupling to matter can be  enhanced by increasing the in-plane electron-hole reduced mass $\mu_{\parallel}$. Moreover, in the absence of doping, we see that the photon completely decouples from the subband and the propagator reduces to the free case $G_c^{(0)}(E)$, since $E_g$ and $E_i$ coincide in the limit $k_F\to0$ [see Eq.~\eqref{eq:EiEg}].

\begin{figure*}[t]
\includegraphics[width=0.9\linewidth]{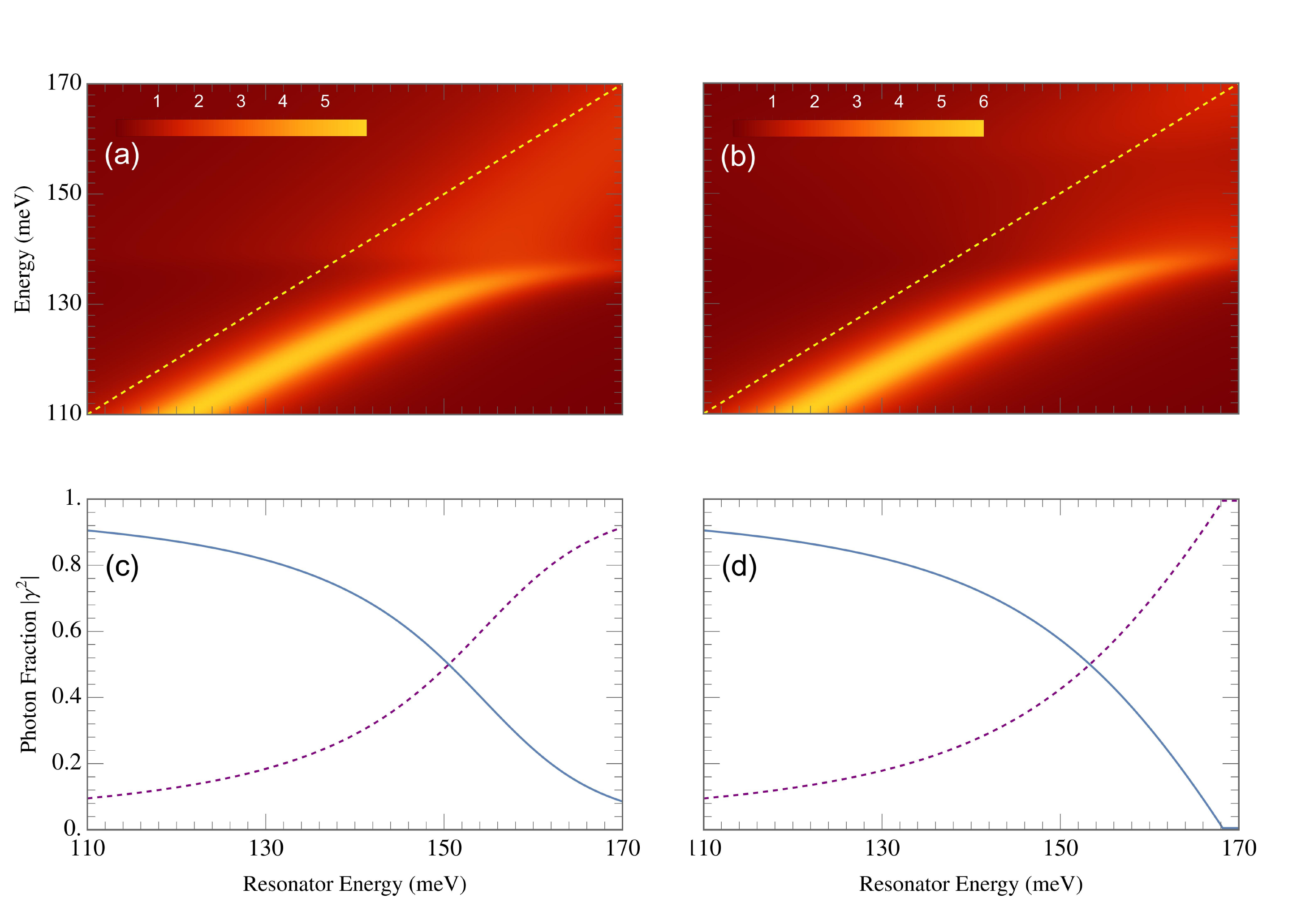}
\caption[system]{\label{SpecPFrac} 
(a,b) Spectral function (in arbitrary units) calculated with (a) and without (b) Coulomb interactions. The dashed yellow line is the bare cavity photon frequency. (c,d) Photon fraction (blue) and matter fraction (purple, dashed) of the exciton polariton for the cases with (c) and without (d) Coulomb interaction. All results are obtained using the physical parameters in Table~\ref{ParameterTable}.
}
\end{figure*}

\begin{figure}
\includegraphics[width=1.0\linewidth]{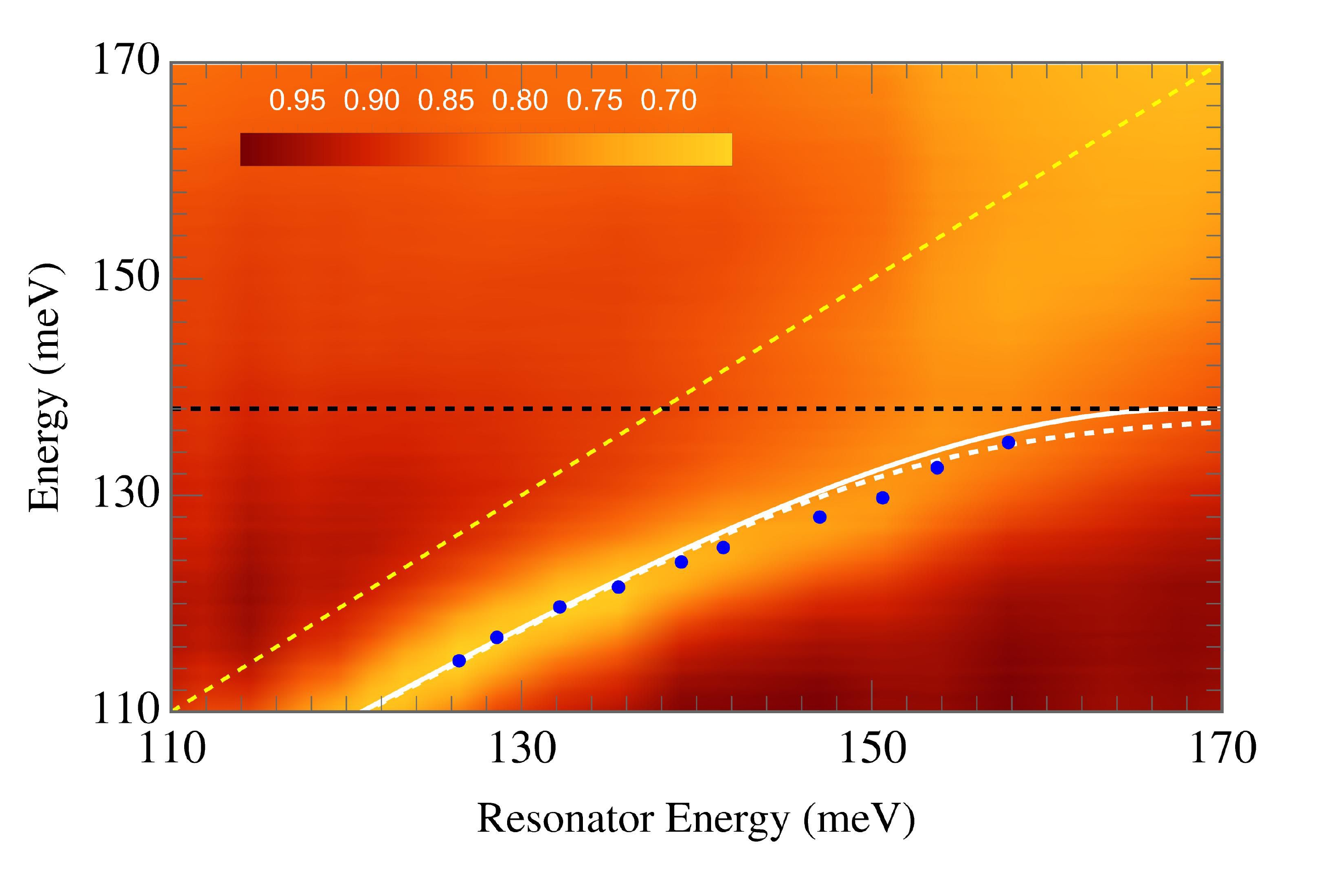}
\caption[system]{\label{fig2} 
The measured reflectance spectrum (in arbitrary units) from the experiment in Ref.~\cite{NatureExBoundByLight}. The onset of the continuum (at 138 meV) and the bare photon energy are displayed as dashed black and dashed yellow lines, respectively. The experimentally extracted peak positions (blue dots) agree well with our predictions for the polariton energy, both with (dashed white) and without (solid white) Coulomb interactions. 
}
\end{figure}

\begin{figure}
\includegraphics[width=1.0\linewidth]{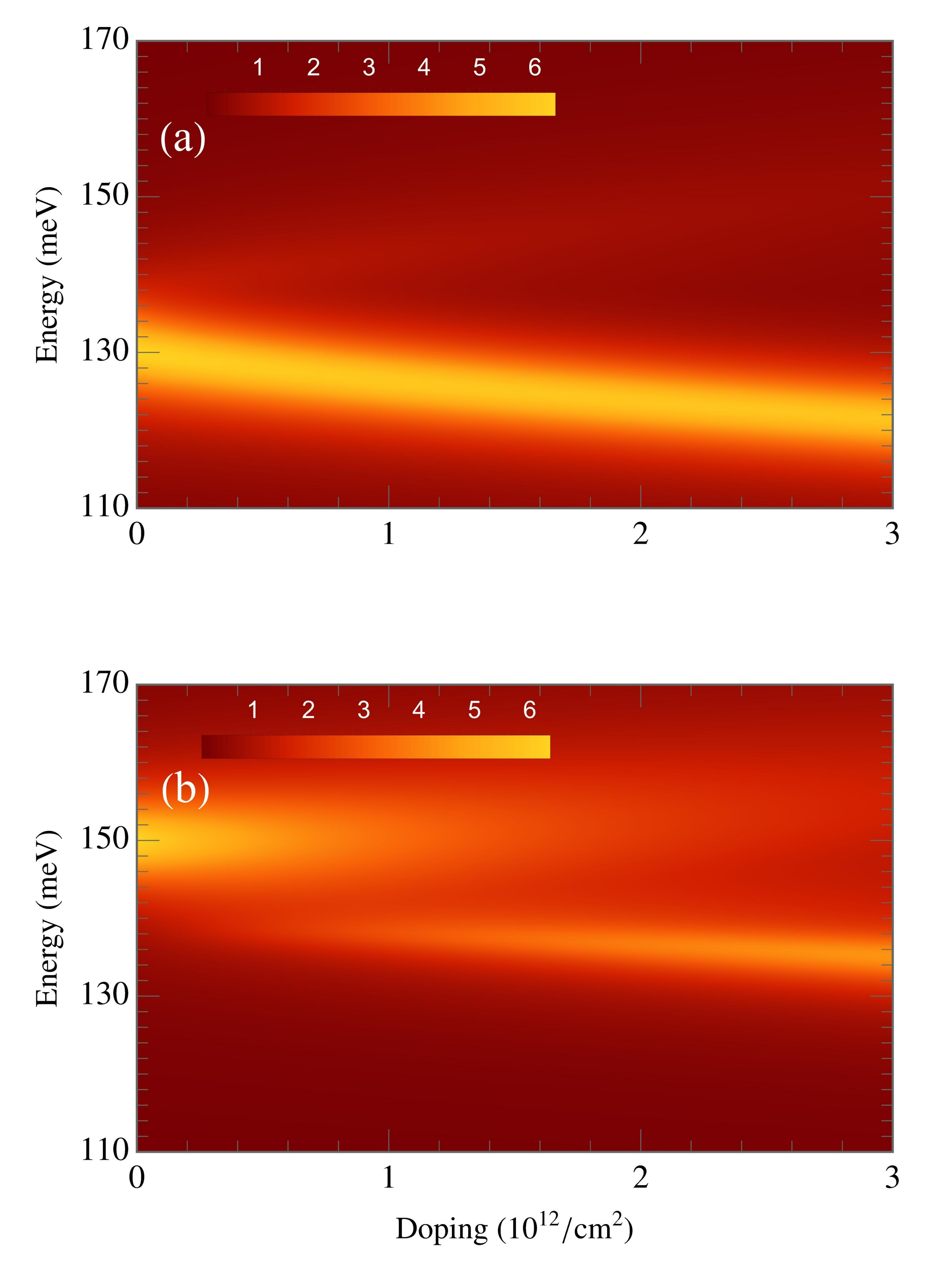}
\caption[system]{\label{DopingSpec} 
 Spectral function (in arbitrary units) versus doping for the resonator energies of 130 meV (a) and 150 meV (b). We show the result of the analytic calculation \eqref{AnalyticProp} where we ignore Coulomb interactions.
}
\end{figure}

\subsection{Results} 
\label{sec:resultsB}

Figure~\ref{SpecPFrac} shows our results for the microcavity spectral function and photon fraction at a fixed electron density in the bound subband, calculated with and without Coulomb interactions. To determine the spectra, we have used the experimental parameters from Ref.~\cite{NatureExBoundByLight}, along with 
estimated parameters for the light-matter coupling strength $g$ and the reduced electron-hole mass $\mu_\parallel$ --- see Table~\ref{ParameterTable}. Below the ionization energy at 138 meV, the spectral function in  Fig.~\ref{SpecPFrac}(a,b) features a prominent discrete peak which is strongly shifted by more than 10 meV from the bare photon resonance. This is due to the existence of a single pole in the dressed photon propagator when $[G_c(E=E_p)]^{-1}=0$ (taking for simplicity $\Gamma=\Gamma_{eh}=0$), corresponding to the energy $E_p$ of the exciton-polariton quasiparticle. 

In Fig.~\ref{fig2} we show the experimentally measured reflectance spectrum from Ref.~\cite{NatureExBoundByLight}. We see that this compares well with our calculated spectral functions in Fig.~\ref{SpecPFrac}. In particular, we find that the position of our exciton-polariton quasiparticle peak below the ionization threshold is in excellent agreement with that extracted from the measured reflectance. 

For resonator frequencies far below the ionization energy, we find that the polariton quasiparticle is almost purely photonic. However, the polariton becomes matter dominated, i.e., excitonic, as the resonator energy approaches the ionization energy. To see this, we note that in the vicinity of the pole, the dressed photon propagator takes the form $G_c(E\simeq E_p)=|\gamma|^2/(E-E_p+i |\gamma|^2\Gamma)$, according to Eq.~\eqref{Discrete G}. Taylor expanding the denominator of the approximation in Eq.~\eqref{AnalyticProp} to first order in $E-E_p$, we obtain the analytic expression for the photon fraction in the absence of Coulomb interactions,
\be \label{APhotonFrac}
|\gamma|^2 = \bigg(
1-\frac{g^2 \mu_\parallel \sqrt{m_{ez}}}{2\sqrt{2} \pi}
\big[(E_g-E_p)^{-\frac{1}{2}}-(E_i-E_p)^{-\frac{1}{2}}\big]
\bigg)^{-1} .
\ee
This explicitly shows that when $E_p$ is far below $E_i$ we have $|\gamma|^2\to1$, whereas when $E_p\to E_i$ we have $|\gamma|^2\to0$. This behavior is illustrated in Fig.~\ref{SpecPFrac}(c,d).

As the polariton approaches the ionization energy, there is a qualitative difference between the calculations with and without Coulomb interactions. Indeed, we find that formally an electron-hole bound state is always present when we include Coulomb interactions, although it is extremely weakly bound for large resonator frequencies, with an associated negligible photon fraction. This is due to the Coulomb potential being effectively attractive along the transverse direction. By contrast, in the absence of Coulomb interactions the exciton polariton enters the ionization continuum at a finite value of the cavity mode energy, as can be seen from the abrupt decrease in the photon fraction to zero in Fig.~\ref{SpecPFrac}(d). To understand this, we set $E_p=E_i$ and solve for the pole of the dressed photon propagator in Eq.~\eqref{AnalyticProp} (taking again $\Gamma=\Gamma_{eh}=0$ for simplicity):
\be \label{Pole}
\w
=E_i+\frac{g^2\sqrt{\mu_\parallel m_{e_z}}k_F}{2\pi}.
\ee
The shift from the value $\omega=E_i$ in the absence of doping can be viewed as a measure of the strength of the light-matter coupling, and we see that this is on the order of 30 meV. 
Importantly, Eq.~\eqref{Pole} shows that the shift is proportional to $k_F$, i.e., it would disappear in the absence of doping, as expected. In spite of this qualitative difference between the two cases, the polariton quasiparticle peaks look qualitatively similar in Fig.~\ref{SpecPFrac}(a,b).

Above the ionization energy, the quasiparticle enters a continuum, and it becomes diffuse. Here, the qualitative difference between the cases with/without Coulomb interactions in Fig.~\ref{SpecPFrac}(a,b) mainly stems from the different estimated values of $\mu_\parallel$, which imply that the energy widths of the 2D continuum, $E_g-E_i$, are different (see Eq.~\eqref{eq:EiEg}). In panel (a), the upper edge of this continuum is not visible, whereas in panel (b) this is at approximately 160 meV (above which we still have a continuum of 3D scattering states).

In Fig.~\ref{DopingSpec}, we instead show the spectral function as a function of electron density, calculated in the analytic case where we ignore the Coulomb interaction. Panels (a) and (b) show the cases where the cavity mode is set below and above the ionization energy, respectively, at 130 meV and 150 meV. In both cases, we can interpret the resulting spectrum as being due to an avoided crossing between the cavity mode and the edge of the 2D continuum. Again we see that the light-matter coupling can be quite large at experimentally realistic electron densities.

Apart from the spectrum, our approach also allows us to directly extract the electron-hole wave function from the solution of the \sch equation~\eqref{CoupledEqns}. In the general case, these must be obtained numerically. However, in the absence of Coulomb interactions, the momentum space wave function is simply obtained from rearranging Eq.~\eqref{Eq1Coupled}. We then obtain analytic position-space expressions in two limits: In the photon dominated regime far away from the continuum where $(E_p-E_i)\gg\frac{k_F^2}{2\mu_\parallel}$, and in the excitonic regime where $E_p\simeq E_i$. 
We write these as $\phi^{(C)}$ and $\phi^{(X)}$ labelling the photon dominated and exciton dominated regimes respectively. 
In the photon-dominated regime, we use the approximation that
$k_F^2/2\mu_\parallel\ll E_i-E_p$ and project onto coordinate space along the in-plane radial direction $r$ and transverse direction $z$ separately to find
\be \label{phiF}
\begin{split}
& \phi^{(C)} (r,0)=\frac{1}{\sqrt{\mathcal{N}}} \frac{g \gamma \sqrt{m_{ez}}}
{2\sqrt{2}\pi \sqrt{E_i-E_p} }
\frac{k_F J_1 (k_F r)}
{r} \\ 
& \phi^{(C)} (0,z)=\frac{1}{\sqrt{\mathcal{N}}} \frac{g \gamma \sqrt{m_{ez}}}
{4\sqrt{2}\pi \sqrt{E_i-E_p} }
k_F^2 e^{-\sqrt{2 m_{ez}(E_i-E_p)} z} .
\end{split}
\ee
Here $J_n$ is the Bessel function of the first kind and the normalization $\mathcal{N}= \frac{(g\gamma)^2 \sqrt{m_{ez}} k_F^2}{8\sqrt{2} \pi (E_i-E_p)^{3/2}}$. Now we proceed in a similar manner for $E_p=E_i$, obtaining just the functional form of the wave function $\phi^{(X)}$ since it is not formally normalizable right at the continuum:
\be \label{phiI}
\begin{split}
& \phi^{(X)} (r,0)= \frac{ {\rm sin}(k_F r)}{k_F r} ,   \\
& \phi^{(X)} (0,z)= \frac{1- e^{-\sqrt{\frac{m_{ez}}{\mu_\parallel}} k_F z }}
{\sqrt{\frac{m_{ez}}{\mu_\parallel}} k_F z} .  
\end{split}
\ee

\begin{figure*}
\includegraphics[width=1.0\linewidth]{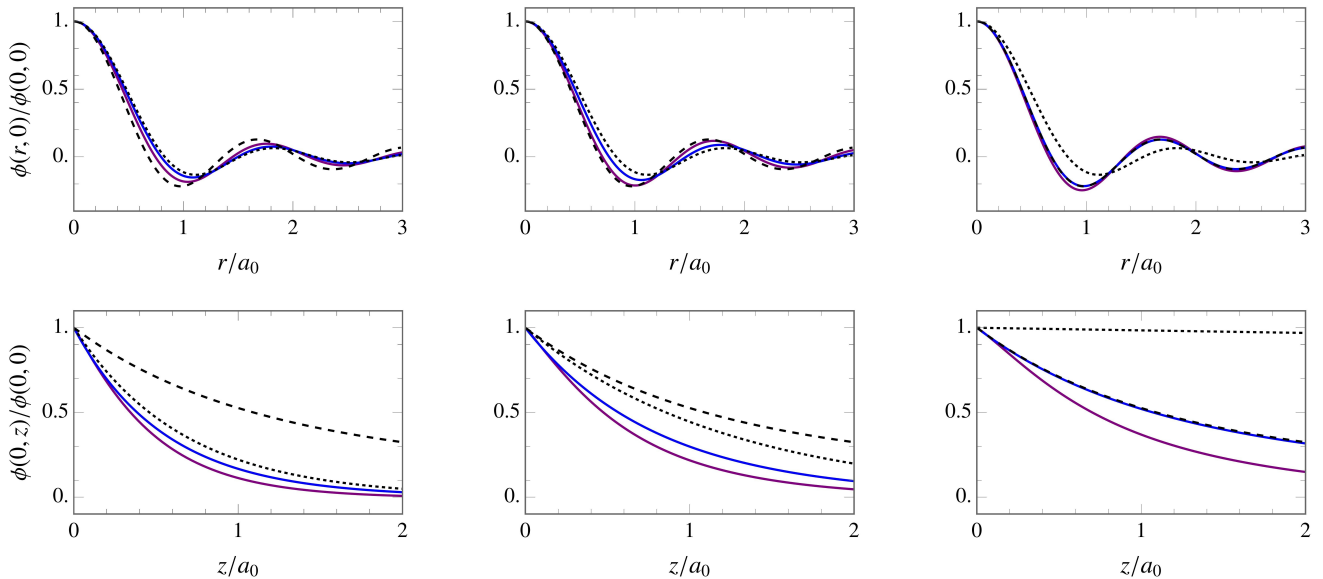}
\caption[system]{\label{fig4} 
Electron-hole wave functions along the in-plane (top) and transverse (bottom) directions. The wave functions with (purple) and without (blue) the Coulomb potential
are shown for resonator energies 130 meV, 150 meV and 168 meV from left to right. We compare these with the analytic approximations $\phi^{(C)}$ (black short dashed) and $\phi^{(X)}$ (black long dashed). To focus on the functional form of the wave functions, we have normalized them by their value at the origin.
}
\end{figure*}

The resulting electron-hole wave functions are shown in Fig.~\ref{fig4}, both as a function of in-plane electron-hole separation, and as a function of transverse separation. Interestingly, in the in-plane direction the wave function is seen to have nodes, which is  different to the usual behaviour expected from a ground state. We attribute this to the negative reduced effective mass of the electron-hole pair, and the presence of the Fermi momentum which means that the Fourier transform is performed on a restricted range of $\vk_\parallel$. Conversely, along the transverse direction, the exciton wave function appears without nodes due to the positive reduced effective mass of the electron-hole pair along this direction. The analytic approximations are seen to match the numerics well in their respective energy ranges; in particular, $\phi^{(C)}$~\eqref{phiF} compares well with the full wave function at lower energies, and the wave function $\phi^{(X)}$~\eqref{phiI} fits better near the continuum.

\section{Conclusion} \label{sec:conc}

To conclude, we have introduced a versatile microscopic formalism to model the scenario of an electron-hole pair that can become bound due to the strong coupling to microcavity photons. 
Using a diagrammatic approach based on Green's functions, we have obtained the spectral function that includes both a discrete exciton-polariton resonance and an unbound electron-hole continuum. Our results compare well with the recent experiment in Ref.~\cite{NatureExBoundByLight} when we use physically realistic parameters for the semiconductor quantum well in the microcavity. Furthermore, by neglecting Coulomb interactions, we were able to obtain analytical results for the spectrum and the electron-hole wave function, which compared well with our numerics for the full problem. In particular, we found that the main effect of the Coulomb potential between electrons and holes was the formation of a weakly bound excitonic state at all resonator energies due to the effective attraction along the transverse direction. This is in contrast to previous theories of excitons bound by photon exchange, which predict a critical value of the resonator energy at which the electron-hole pair unbinds~\cite{NatureExBoundByLight,StrongCouplingOfIonTrans}.

The exciton polaritons investigated in this work carry a formal similarity to polaron-polaritons~\cite{Sidler2017}. In the latter case, an exciton polariton is coherently dressed by particle-hole excitations of a Fermi sea~\cite{Sidler2017,Efimkin2017}, leading to the formation of attractive and repulsive polariton quasiparticles. By analogy, in the present scenario the discrete resonance below the ionization energy may be viewed as an attractive photon polaron, while the resonance in the continuum may be viewed as a metastable repulsive photon polaron~\cite{Scazza2022}. In the future, it would be interesting to explore this analogy further, for instance to investigate the nature of interactions between such photon polarons due to their electronic constituents~\cite{Tan2020}. In particular, the interactions between conventional exciton polaritons have been shown to depend strongly upon the light-matter coupling strength~\cite{Bleu2020PRR}, and likewise the interactions between photon polarons would be expected to depend strongly and non-trivially on the quantum well doping.

\acknowledgements
This work was supported by the Australian Research Council Center of Excellence in Future Low-Energy Electronics Technologies (CE170100039). M.M.P. and J.L. were supported through the Australian Research Council Future Fellowships FT200100619 and FT160100244, respectively.

\appendix

\section{Derivation of photon propagator} \label{AppendixA}
We start by defining the Green's operator
\be
\hat{G}(E)=\frac{1}{E^+-\hat{H}_p-\hat{H}_m-\hat{H}_g},
\ee
where we have introduced an infinitesimal imaginary part such that $E^+=E+i0$.
Expanding in the light-matter coupling term $\hat{H}_g$ gives the infinite series
\be
\begin{split} \label{eq:Gseries}
&\hat{G}(E)=\frac{1}{E^+-\hat{H}_p-\hat{H}_m} + \\ & 
\frac{1}{E^+-\hat{H}_p-\hat{H}_m} \hat{H}_g
\frac{1}{E^+-\hat{H}_p-\hat{H}_m}
+...
\end{split}
\ee
To obtain the photon propagator, we must take the expectation value of the Green's operator with the cavity photon state, yielding $G_c(E) = \bra{0} \hat{c} \, \hat{G}(E) \hat{c}^\dag\ket{0}$. The term $\hat{H}_g$ in Eq.~\eqref{eq:Gseries} is responsible for transforming the photon state to an electron-hole state and vice versa. Evaluating odd powers of $\hat{H}_g$ between photon states results in an inner product between a photon state and an electron-hole state, which evaluates to zero, leaving only the even terms in the expansion. Without explicitly evaluating the photon states, we therefore write the photon propagator in its operator form as
\be
\begin{split}  \label{eq:Gseries2}
&\hat{G}_c (E)=\frac{1}{E^+-\hat{H}_p
} + \\ & 
\frac{1}{E^+-\hat{H}_p
} \hat{H}_g
\frac{1}{E^+
-\hat{H}_m}
\hat{H}_g
\frac{1}{E^+-\hat{H}_p
}
+...
\end{split}
\ee
Here, we have used the fact that the matter term $\hat{H}_m$ evaluates to zero when the photon state is applied to the left- and right-most terms. In addition, the $\hat{H}_g$ term transforms the photon into an electron-hole state, which in turn allows us to drop the photon term $\hat{H}_p$ in the denominator of the middle propagator in the second line of Eq.~\eqref{eq:Gseries2}. This results in alternating $\hat{H}_p$ and $\hat{H}_m$ terms in the denominators within the series. Writing $\frac{1}{E^+-\hat{H}_p}$ as the free photon propagator $\hat{G}_c^{(0)} (E)$ and $\frac{1}{E^+-\hat{H}_m}$ as the full electron-hole propagator $\hat{G}_m (E)$ (which includes the Coulomb potential), the dressed photon propagator in operator form is
\be \label{eq:photonprop}
\begin{split}
&\hat{G}_c (E) =
\hat{G}_c^{(0)} (E) + 
\hat{G}_c^{(0)} (E) 
\hat{H}_g
\hat{G}_m (E)
\hat{H}_g
\hat{G}_c^{(0)} (E) + \\
&\hat{G}_c^{(0)} (E)
\hat{H}_g
\hat{G}_m (E)
\hat{H}_g
\hat{G}_c^{(0)} (E)
\hat{H}_g
\hat{G}_m (E)
\hat{H}_g
\hat{G}_c^{(0)} (E) + ...
\end{split}
\ee

The infinite expansion in Eq.~\eqref{eq:photonprop} can be resummed to give the final equation 
\be \label{PropEqn}
\begin{split}
\hat{G}_c (E)
&=\frac{1}{\hat{G}_c^{(0)} (E)^{-1} -\hat{H}_g \hat{G}_m (E) \hat{H}_g } \\
&=\frac{1}{\hat{G}_c^{(0)} (E)^{-1} - \hat{\Sigma}(E) },
\end{split}
\ee
where the term $\hat{H}_g \hat{G}_m (E) \hat{H}_g$ is the self energy $\hat{\Sigma}(E)$. This corresponds to the Dyson equation~\cite{Fetter} of $G_c(E)$
\be \label{Dyson}
\hat{G}_c (E) = \hat{G}_c^{(0)} (E) + 
\hat{G}_c^{(0)} (E) \hat{\Sigma} (E) \hat{G}_c (E),
\ee
which we illustrate in Fig.~\ref{DysonDiagram}. 

\bibliography{references}

\end{document}